\documentclass[journal=inoraj,manuscript=article,layout=twocolumn]{achemso}
\pdfoutput=1
\usepackage{chemformula}
\usepackage{gensymb}
\usepackage{graphicx}
\usepackage{booktabs}
\usepackage{multirow}
\usepackage{longtable}

\newcommand*\LnCOB{Ca$_4$\emph{Ln}O(BO$_3$)$_3$}
\newcommand*\LaCOB{Ca$_4$LaO(BO$_3$)$_3$}
\newcommand*\PrCOB{Ca$_4$PrO(BO$_3$)$_3$}

\newcommand*\GdCOB{Ca$_4$GdO(BO$_3$)$_3$}
\newcommand*\TbCOB{Ca$_4$TbO(BO$_3$)$_3$}
\newcommand*\DyCOB{Ca$_4$DyO(BO$_3$)$_3$}
\newcommand*\HoCOB{Ca$_4$HoO(BO$_3$)$_3$}

\newcommand*\YbCOB{Ca$_4$YbO(BO$_3$)$_3$}

\author{Nicola D. Kelly}
\email{ne281@cam.ac.uk}
\author{Si\^{a}n E. Dutton}
\email{sed33@cam.ac.uk}
\affiliation[University of Cambridge]
{Cavendish Laboratory, University of Cambridge, J J Thomson Avenue, Cambridge, CB3~0HE, UK}
\title[Magnetic properties of lanthanide calcium oxyborates]{Magnetic properties of quasi-one-dimensional lanthanide calcium oxyborates \LnCOB}

\begin{document}
\begin{abstract}
This study examines the lanthanide calcium oxyborates \LnCOB\ (\emph{Ln}~= La, Pr, Nd, Sm, Eu, Gd, Tb, Dy, Ho, Y, Er, Yb). The reported monoclinic structure (space group \emph{Cm}) was confirmed using powder X-ray diffraction. The magnetic \emph{Ln}$^{3+}$ ions are situated in well-separated chains parallel to the $c$ axis in a quasi-one-dimensional array. Here we report the first bulk magnetic characterisation of \LnCOB\ using magnetic susceptibility $\chi(T)$ and isothermal magnetisation $M(H)$ measurements at $T \geq 2$~K. With the sole exception of \TbCOB, which displays a transition at $T$ = 3.6~K, no magnetic transitions occur above 2~K, and Curie-Weiss analysis indicates antiferromagnetic nearest-neighbour interactions for all samples. Calculation of the magnetic entropy change $\Delta S_m$ indicates that \GdCOB\ and \HoCOB\ are viable magnetocaloric materials at liquid helium temperatures in the high-field and low-field regimes respectively.
\end{abstract}

\section{Introduction}
An ideal one-dimensional (1D) system -- a single chain of magnetic ions -- would never display long-range order \cite{Mermin1966,Blundell2007}. Such systems have been predicted to host exotic magnetic behaviour such as spinons \cite{Haldane1991}, but are impossible to realise in the solid state. Certain crystal structures may, however, be quasi-1D if the spin chains are kept well separated by nonmagnetic atoms, but in most cases there is weak coupling between the chains, leading to 3D ordering at low temperatures. Extensive work has been carried out on quasi-1D $S=\frac{1}{2}$ systems containing first-row transition metals, but much less is known about lanthanide systems \cite{Vasiliev2018,Glawion2011,Mourigal2012,Dutton2012,Hase1993}. These have very different magnetic properties from the 3$d$ transition metals as a result of strong spin-orbit coupling, and the interplay between superexchange ($J$), dipolar ($D$) and crystal electric field (CEF) effects means that they are likely to require different models for the magnetism. Examples of quasi-1D lanthanide systems include the hydroxycarbonates \emph{Ln}OHCO$_3$ and formates \emph{Ln}(HCOO)$_3$. Both series of compounds have been shown to display unusual magnetic properties and to be viable magnetocaloric materials at temperatures below 10~K, although in some cases inter-chain coupling produces three-dimensional ordering at temperatures below 2~K \cite{Harcombe2016,Dixey2018,Dixey2019}.

Low-dimensional magnetic systems are of interest for solid-state magnetic refrigeration as a more sustainable alternative to liquid helium for cooling to low temperatures. This technology relies on the magnetocaloric effect (MCE) arising from adiabatic demagnetisation of the sample: this causes the temperature to drop as magnetic domains de-align from the field direction, enabling large amounts of magnetic entropy to be extracted. The lower limit for magnetic cooling is set by the long-range magnetic ordering temperature of the sample; thus, low-dimensional materials are useful as they usually display suppression of this ordering temperature. Lanthanide compounds have been widely studied for this purpose, particularly those containing Gd$^{3+}$, which has a large spin compared to the transition metals, and usually\cite{Paddison2015} no crystal field interactions compared with the other lanthanides, since $L=0$ \cite{Barclay1982,Numazawa2003,Dixey2018,Harcombe2016}. For Gd$^{3+}$ (e.g.~Gd$_3$Ga$_5$O$_{12}$, `GGG') the MCE is maximised in high fields ($\mu_0H > 5$~T), where Heisenberg systems have been shown to be optimised, but obtaining such fields still requires liquid helium to cool the superconducting magnets. At low fields $\mu_0H \leq 2$~T a permanent magnet can be used to provide the external field; systems with significant single-ion anisotropy (e.g.~Dy$_3$Ga$_5$O$_{12}$) tend to perform better in this regime \cite{Numazawa2003,Saines2015}.

\begin{figure}
\includegraphics[width=8cm]{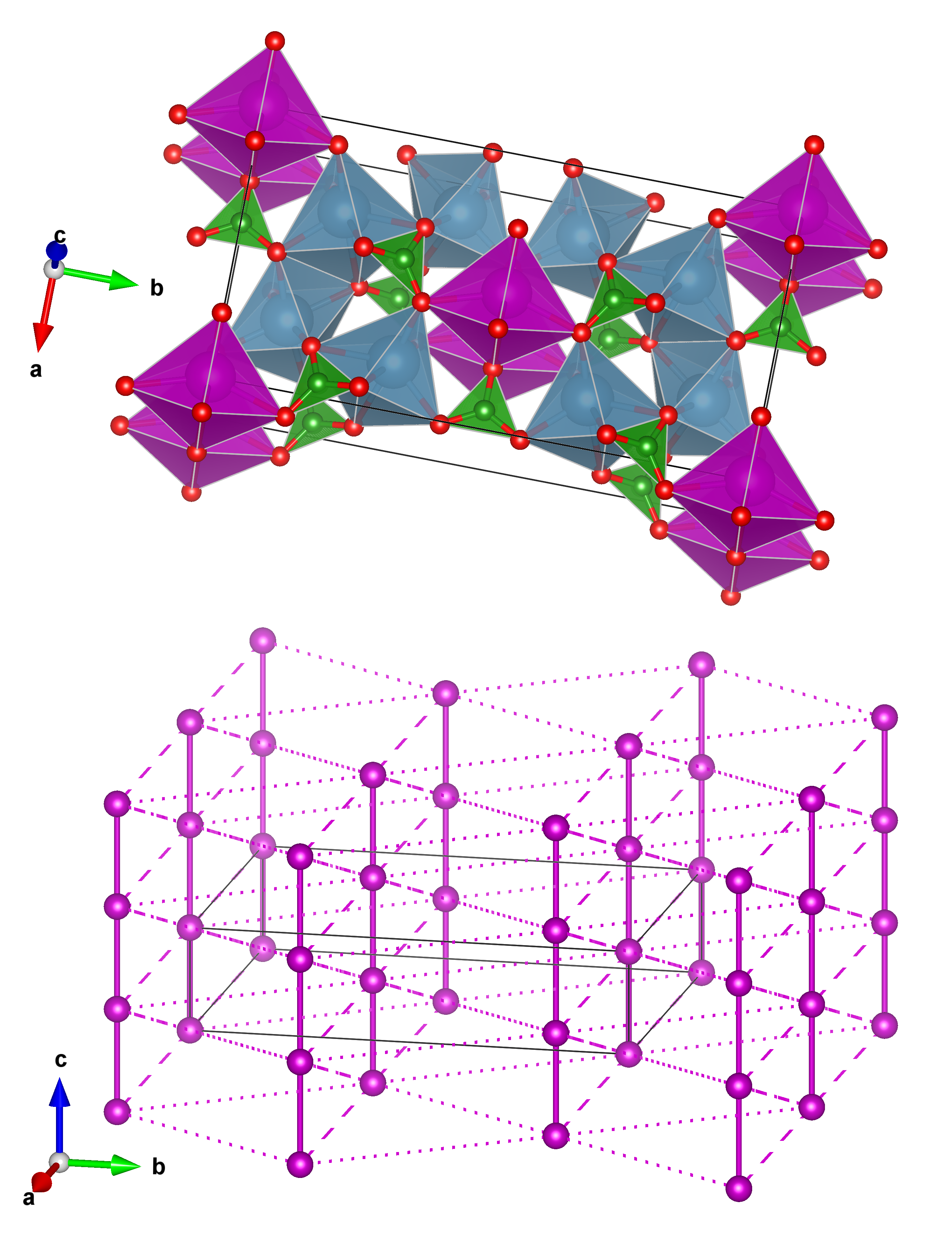} 
\caption{Top: Monoclinic crystal structure of \LnCOB\ (\emph{Ln}~=~Y, La--Lu). O atoms in red; CaO$_6$ and \emph{Ln}O$_6$ distorted octahedra in blue and purple respectively; trigonal planar (BO$_3$)$^{3-}$ groups in green. Bottom: Connectivity of \emph{Ln}$^{3+}$ ions in \LnCOB: intra-chain (solid lines, 3.6--3.8~\AA) and inter-chain (dashed lines, 8.1--8.3~\AA, and dotted lines, 9.0--9.2~\AA).}
\label{fig:lncobstructure}
\end{figure}

The lanthanide calcium oxyborates with general formula \LnCOB\ (\emph{Ln}~= Y, La--Lu) have previously been investigated as nonlinear optic materials \cite{Reuther2011,Mockel2013}. The synthesis of these compounds in powder form may be carried out through straightforward solid-state or sol-gel procedures \cite{Norrestam1992,Zhang2005}. The lanthanide ions in the unit cell are arranged in chains parallel to the $c$-axis (Fig.~\ref{fig:lncobstructure}) with ions separated by $c \approx$ 3.6 \AA\ along the chains. These chains are well separated in the $ab$ plane by 8--9 \AA, leading to a quasi-1D \emph{Ln}$^{3+}$ array.

In this article we report the solid-state synthesis of twelve compounds with the formula \LnCOB\ from across the lanthanide series, followed by structural characterisation using powder X-ray diffraction (PXRD). Furthermore we report the bulk magnetic characterisation of these compounds using magnetic susceptibility and isothermal magnetisation measurements at $T \geq 2$~K. Their potential application in magnetic refrigeration applications is discussed with the aid of magnetic entropy calculations.

\section{Experimental}

Polycrystalline samples of \LnCOB\ (\emph{Ln}~= La, Pr, Nd, Sm, Eu, Gd, Tb, Dy, Ho, Y, Er, Yb) were synthesised according to a ceramic procedure, adapted from Ref.~\citenum{Crossno1997}, from \ch{CaCO3} (99.99~\%), \ch{H3BO3} (99.999~\%) and \emph{Ln}$_2$O$_3$ (\emph{Ln}~= La, Nd, Sm, Eu, Gd, Dy, Ho, Y, Er, Yb), \ch{Pr6O11} or \ch{Tb4O7} (all lanthanide oxides $\geq$~99.99~\%). Lanthanide oxides were pre-dried at 800~\degree C overnight prior to weighing out. Stoichiometric amounts of the reagents (except in the case of \emph{Ln}~= Yb: see below) were ground with a pestle and mortar and placed in an alumina crucible. The powder was first heated in air at 900~\degree C for 4~h in order to effect decomposition of the boric acid and calcium carbonate. The sample was subsequently cooled, reground and reheated to 1200~\degree C for several days, with intermediate regrinding every 24 hours, until the percentages of impurity phases no longer changed.

Room temperature PXRD patterns were collected on a Bruker D8 diffractometer (Cu K$\alpha$, $\lambda = 1.541$ \AA) in the range $10 \leq 2\theta(\degree) \leq 90$ with a step size of 0.01\degree\ and measurement time 1 second per step. Rietveld refinement\cite{Rietveld1969} was carried out using the program Topas.\cite{Coelho2018}

Magnetic susceptibility and isothermal magnetisation were measured on a Quantum Design 9 T Physical Properties Measurement System using the ACMS-II option in the temperature and field ranges $2 \leq T(K) \leq 300$ and $0 \leq \mu_0H(T) \leq 9$ respectively. In a low field of 500 Oe, the $M(H)$ curve is linear for all $T$ and the susceptibility can therefore be approximated by $\chi(T)=M/H$.

\section{Results}
\subsection{Crystal Structure}
\begin{figure}
\includegraphics[width=8cm]{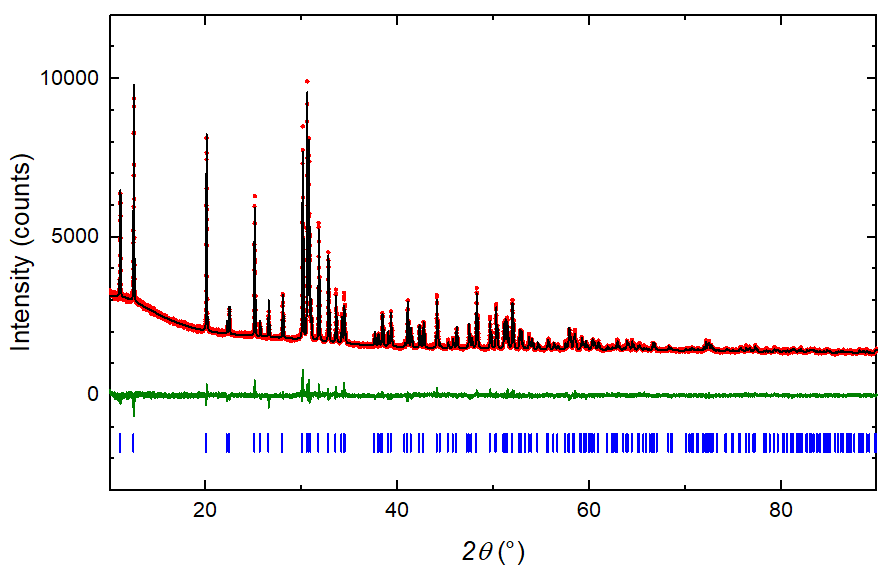}
\caption{Room temperature PXRD pattern for \DyCOB: Red dots -- experimental data; black line -- calculated intensities; green line -- difference pattern; blue tick marks -- Bragg reflection positions.}
\label{fig:rtpxrddycob}
\end{figure}

From PXRD and Rietveld refinement all samples were found to adopt the previously reported monoclinic \emph{Cm} structure (Fig.~\ref{fig:lncobstructure}); a representative X-ray refinement is given in Fig.~\ref{fig:rtpxrddycob} \cite{Norrestam1992,Ilyukhin1993a,Crossno1997}. The unit cell contains two independent Ca$^{2+}$ sites, both six-coordinate but distorted from perfectly octahedral geometry. Additionally the crystal structure contains distorted \emph{Ln}O$_6$ octahedra, which share edges along the `chains', and trigonal planar BO$_3^{3-}$ groups, which are tilted at different angles from the $c$-axis. A previous study found that the tilting of each of these borate groups changed in a regular way as the size of the lanthanide ions was varied \cite{Crossno1997}, but such analysis is beyond the scope of this work, which is limited by the powder samples and solely X-ray diffraction measurements.

\begin{figure} 
\centering
\includegraphics[width=8cm]{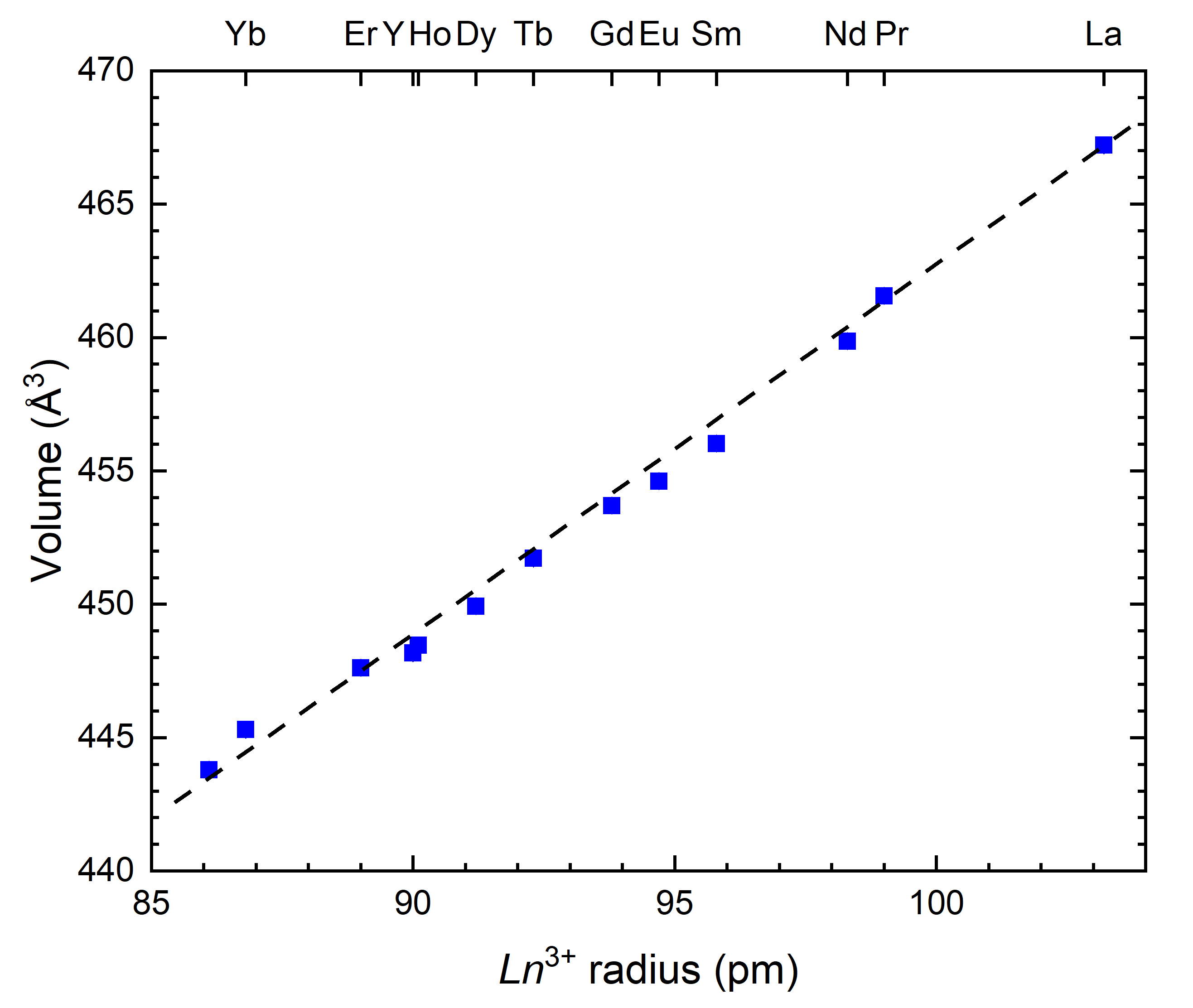}
\caption{Unit cell volume of all \LnCOB\ samples as a function of lanthanide ionic radius. Error bars (from individual refinements) are smaller than the datapoints. The dashed line is given as a guide to the eye.}
\label{fig:lncobvolume}
\end{figure}

In some cases small amounts, $\leq 5$ wt \%, of non-magnetic impurities (\ch{Ca3(BO3)2}, \ch{H3BO3}) remained after multiple heating steps: for details, see Table~\ref{table:lncobparamspxrd}. For \YbCOB, 50~\%\ excess \ch{H3BO3} was required to ensure complete reaction of \ch{Yb2O3}, leading to a higher proportion of \ch{Ca3(BO3)2} in the final sample. There is also a larger proportion of \ch{Ca3(BO3)2} in \LaCOB\ than in the other samples; this is attributed to either incomplete drying or reabsorption of water into \ch{La2O3} before weighing out. The series \LnCOB\ obeys Vegard's Law for variation of lattice parameters (Fig.~\ref{fig:lncobvolume}).

The insensitivity of X-ray diffraction to oxygen and boron in the presence of heavy elements (Ca, \emph{Ln}) required the atomic positions of O and B to be fixed at previously reported values \cite{Ilyukhin1993}, while the atomic coordinates of Ca were refined (Table~\ref{table:lncobparamspxrd}). Considering ionic radii, the trivalent lanthanide ions range in size from 103.2 pm (La) down to 86.8 pm (Yb) while 6-coordinate Ca$^{2+}$ is 100 pm \cite{Shannon1976}; the extent of cation mixing might then be expected to increase with increasing \emph{Ln}$^{3+}$ radius. The presence, if any, of Ca$^{2+}$/\emph{Ln}$^{3+}$ site disorder was tested by setting a suitable mixed Ca$^{2+}$ and \emph{Ln}$^{3+}$ occupancy in each of the three metal sites and refining with the overall ratio fixed at 4:1. No significant site disorder was observed for any compound, however, in agreement with a previous single-crystal study of \LaCOB\ \cite{Reuther2011}. The fractional occupancies were therefore fixed at 1 for each site.

\begin{table*} 
\caption{Refined crystal structure parameters for \LnCOB\ samples, from room-temperature PXRD refinements in space group \emph{Cm}. Due to the low scattering power of B and O compared with Ca and \emph{Ln}, the boron and oxygen atomic positions were kept fixed at the following general sites as given for \GdCOB\ \cite{Ilyukhin1993}: B1 (2$a$) = (0.3764, 0, 0.7011); B2 (4$b$) = (0.9491, 0.1947, 0.0798); O1 (2$a$) = (0.8252, 0, 0.4175); O2 (4$b$) = (0.4614, 0.9257, 0.7492); O3 (2$a$) = (0.2032, 0, 0.6043); O4 (4$b$) = (0.0859, 0.1434, 0.0766); O5 (4$b$) = (0.9675, 0.2695, 0.2746). The lanthanide ion is at site 2$a$ = (0, 0, 0). The values of thermal parameters for all atoms were kept fixed at $B_{iso}$ = 1 \AA$^2$.}
\label{table:lncobparamspxrd}
\resizebox{\textwidth}{!}{  
\begin{tabular}{c c c c c c c c}
\toprule
\emph{Ln} & & La & Pr & Nd & Sm & Eu & Gd \\ 
\midrule
\emph{a} (\AA) & & 8.16647(12) & 8.13495(12) & 8.13043(10) & 8.10564(8) & 8.10477(13) & 8.10501(9) \\ 
\emph{b} (\AA) & & 16.07609(24) & 16.06719(19) & 16.05569(16) & 16.03834(15) & 16.04177(21) & 16.03210(15) \\
\emph{c} (\AA) & & 3.63064(5) & 3.60222(4) & 3.59307(3) & 3.57830(3) & 3.56572(4) & 3.56012(3) \\
$\beta$ (\degree) & & 101.4244(8) & 101.3877(7) & 101.3497(6) & 101.3625(5) & 101.3000(8) & 101.2624(6) \\
Volume (\AA$^3$) & & 467.204(11) & 461.562(10) & 459.865(8) & 456.065(7) & 454.610(11) & 453.694(8) \\
$\chi^2$ & & 2.14 & 2.12 & 2.84 & 1.42 & 2.17 & 1.88 \\
$R_{wp}$ & & 6.23 & 5.76 & 6.72 & 4.55 & 5.61 & 5.06 \\
\ch{H3BO3} wt \%\ & & -- & -- & 3.7(2) & 4.0(2) & -- & -- \\
\ch{Ca3(BO3)2} wt \%\ & & 14.2(4) & -- & -- & -- & -- & 2.1(2) \\
Ca1: 4$b$ & $x$ & 0.1479(6) & 0.1420(6) & 0.1402(5) & 0.1424(5) & 0.1352(7) & 0.1377(5) \\
& $y$ & 0.3864(3) & 0.3863(2) & 0.3865(2) & 0.3867(2) & 0.3884(3) & 0.3872(2) \\
& $z$ & 0.3288(18) & 0.3300(14) & 0.3280(13) & 0.3274(13) & 0.3245(17) & 0.3259(14) \\
Ca2: 4$b$ & $x$ & 0.2727(6) & 0.2699(5) & 0.2679(5) & 0.2675(5) & 0.2645(6) & 0.2677(5) \\
& $y$ & 0.1801(3) & 0.1801(3) & 0.1804(3) & 0.1801(2) & 0.1796(3) & 0.1791(3) \\
& $z$ & 0.6813(15) & 0.6691(12) & 0.6657(12) & 0.6560(11) & 0.6618(14) & 0.6591(11) \\
\midrule
\emph{Ln} & & Tb & Dy & Ho & Y & Er & Yb \\
\midrule
\emph{a} (\AA) & & 8.08902(8) & 8.08181(8) & 8.07967(9) & 8.07438(18) & 8.07763(13) & 8.06071(14) \\
\emph{b} (\AA) & & 16.03226(13) & 16.02670(14) & 16.01419(15) & 16.0144(3) & 16.01138(22) & 16.00508(25) \\
\emph{c} (\AA) & & 3.55129(3) & 3.54124(3) & 3.53306(3) & 3.53071(6) & 3.52770(4) & 3.51785(5) \\
$\beta$ (\degree) & & 101.2383(5) & 101.2082(5) & 101.1829(5) & 101.1927(12) & 101.1716(8) & 101.1351(9) \\
Volume (\AA$^3$) & & 451.719(6) & 449.930(7) & 448.461(7) & 447.861(15) & 447.606(11) & 445.302(12) \\ 
$\chi^2$ & & 2.29 & 1.84 & 2.97 & 3.80 & 4.12 & 3.78 \\
$R_{wp}$ & & 5.53 & 4.82 & 5.81 & 7.95 & 6.32 & 9.56 \\
\ch{Ca3(BO3)2} wt \%\ & & -- & -- & 1.5(2) & 5.1(3) & 4.1(2) & 43.7(2) \\
Ca1: 4$b$ & $x$ & 0.1348(5) & 0.1342(5) & 0.1371(4) & 0.1345(10) & 0.1337(6) & 0.1355(10) \\ 
& $y$ & 0.3894(2) & 0.3882(2) & 0.3890(2) & 0.3868(4) & 0.3903(2) & 0.3910(5) \\ 
& $z$ & 0.3184(13) & 0.3149(13) & 0.3163(11) & 0.322(3) & 0.3179(16) & 0.291(3) \\ 
Ca2: 4$b$ & $x$ & 0.2627(4) & 0.2645(4) & 0.2608(4) & 0.2617(9) & 0.2631(5) & 0.2547(10) \\
& $y$ & 0.1781(2) & 0.1791(2) & 0.1784(2) & 0.1768(5) & 0.1746(3) & 0.1759(6) \\ 
& $z$ & 0.6559(11) & 0.6556(10) & 0.6526(9) & 0.6499(21) & 0.6559(13) & 0.6471(24) \\ 
\bottomrule
\end{tabular}}
\end{table*}

\subsection{Bulk magnetic properties}
\begin{figure*} 
\centering
\includegraphics[width=18cm]{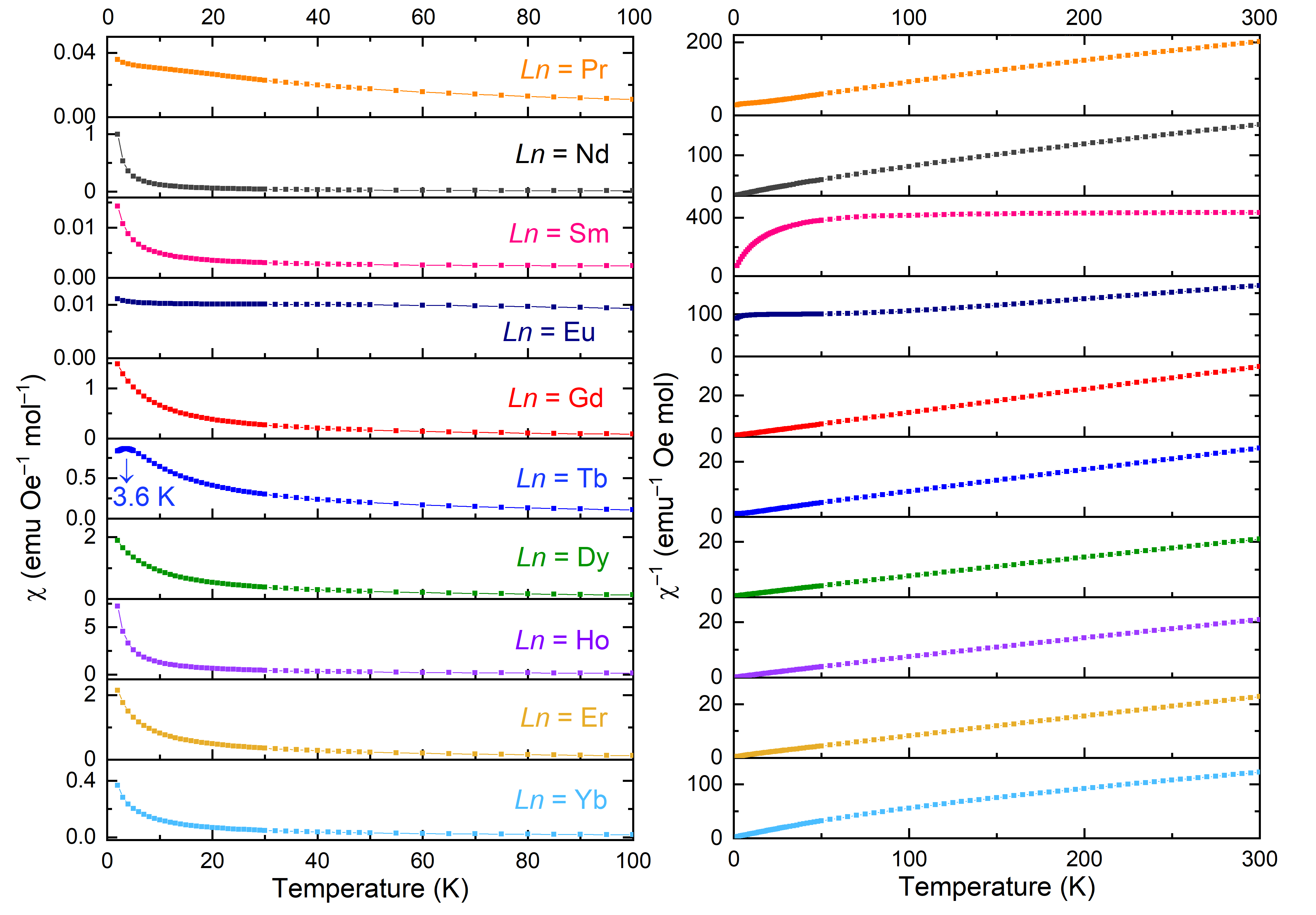}
\caption{Left: Magnetic susceptibility as a function of temperature for the \LnCOB~samples with \emph{Ln}~= Pr, Nd, Sm, Eu, Gd, Tb, Dy, Ho, Er and Yb. Right: Reciprocal magnetic susceptibility, $\chi^{-1}$.}
\label{fig:lncobmvtcombined}
\end{figure*}

The zero-field-cooled magnetic susceptibility curves, collected on warming at $H$~= 500 Oe, are shown in Fig.~\ref{fig:lncobmvtcombined} for \emph{Ln}~= Pr, Nd, Sm, Eu, Gd, Tb, Dy, Ho, Er and Yb. The samples containing Sm and Eu do not obey the Curie-Weiss law ($\chi = \frac{C}{T-\theta_{CW}}$), but display van Vleck paramagnetism due to the mixing of ground states with low-lying excited states \cite{Blundell2001}. The broad feature at $T > $ 6~K for \PrCOB\ is attributed to a singlet ground state with van Vleck paramagnetism, in accordance with previous reports of Pr compounds \cite{Zorko2010,Dun2017a,Mukherjee2017}. A broad magnetic transition characteristic of low-dimensional ordering is visible at $T$ = 3.6 K for \emph{Ln}~= Tb, while for all other \LnCOB\ no transition occurs at $T \geq$ 2~K. Linear Curie-Weiss fitting was carried out in both high-temperature (50--150~K) and low-temperature regimes. The susceptibility of lanthanide compounds is strongly dependent on crystal electric field effects and the low-temperature fitting ranges were therefore varied depending on the lanthanide ion in question \cite{Ashtar2019,Guo2019}. The resultant magnetic parameters are given in Table~\ref{table:lncobmagnetic}. All Curie-Weiss temperatures $\theta_{CW}$ are negative, indicating antiferromagnetic interactions. The effective magnetic moment per \emph{Ln}$^{3+}$ ion was calculated from the Curie constant $C$ for each compound and the high $T$ values agree well with the theoretical free-ion value $g_J\sqrt{J(J+1)}$ within the bounds of experimental error. The Curie constants and thus the effective magnetic moments are broadly consistent between the two fitting regimes.

\begin{table*} 
\caption{Bulk magnetic properties of \LnCOB, \emph{Ln}~= Pr, Nd, Gd--Er, Yb. }
\label{table:lncobmagnetic}
\resizebox{\textwidth}{!}{  
\begin{tabular}{c c c c c c c c}
\toprule
\emph{Ln} & Free-ion moment ($\mu_B$) & High $T$ fit (K) & $\mu_{eff}$ ($\mu_B$) & $\theta_{CW}$ (K) & Low $T$ fit (K) & $\mu_{eff}$ ($\mu_B$) & $\theta_{CW}$ (K) \\
\midrule
Pr & 3.58 & 50--150 & 3.50(7) & --39.2(8) & 20--50 & 3.48(7) & --37(1) \\
Nd & 3.62 & 50--150 & 3.60(7) & --15.9(3) & 20--50 & 3.28(7) & --3.07(6) \\
Gd & 7.94 & 50--150 & 8.41(17)& --3.40(7) & 2--25 & 8.50(17) & --3.85(8) \\
Tb & 9.72 & 50--150  & 10.0(2) & --13.9(3) & 25--50 & 9.57(19) & --8.25(17) \\
Dy & 10.65 & 50--150 & 10.8(2) & --9.8(2) & 15--30 & 10.3(2) & --4.78(10) \\ 
Ho & 10.61 & 50--150 & 10.7(2) & --5.68(11) & 10--50 & 10.3(2) & --0.657(13) \\
Er & 9.58 & 50--150 & 10.3(2) & --8.20(16) & 2--25 & 9.64(19) & --3.99(8) \\
Yb & 4.54 & 50--150 & 4.28(9) & --26.4(5) & 2--20 & 3.50(7) & --2.52(5)\\
\bottomrule
\end{tabular}}
\end{table*}

\begin{figure}
\includegraphics[width=8.5cm]{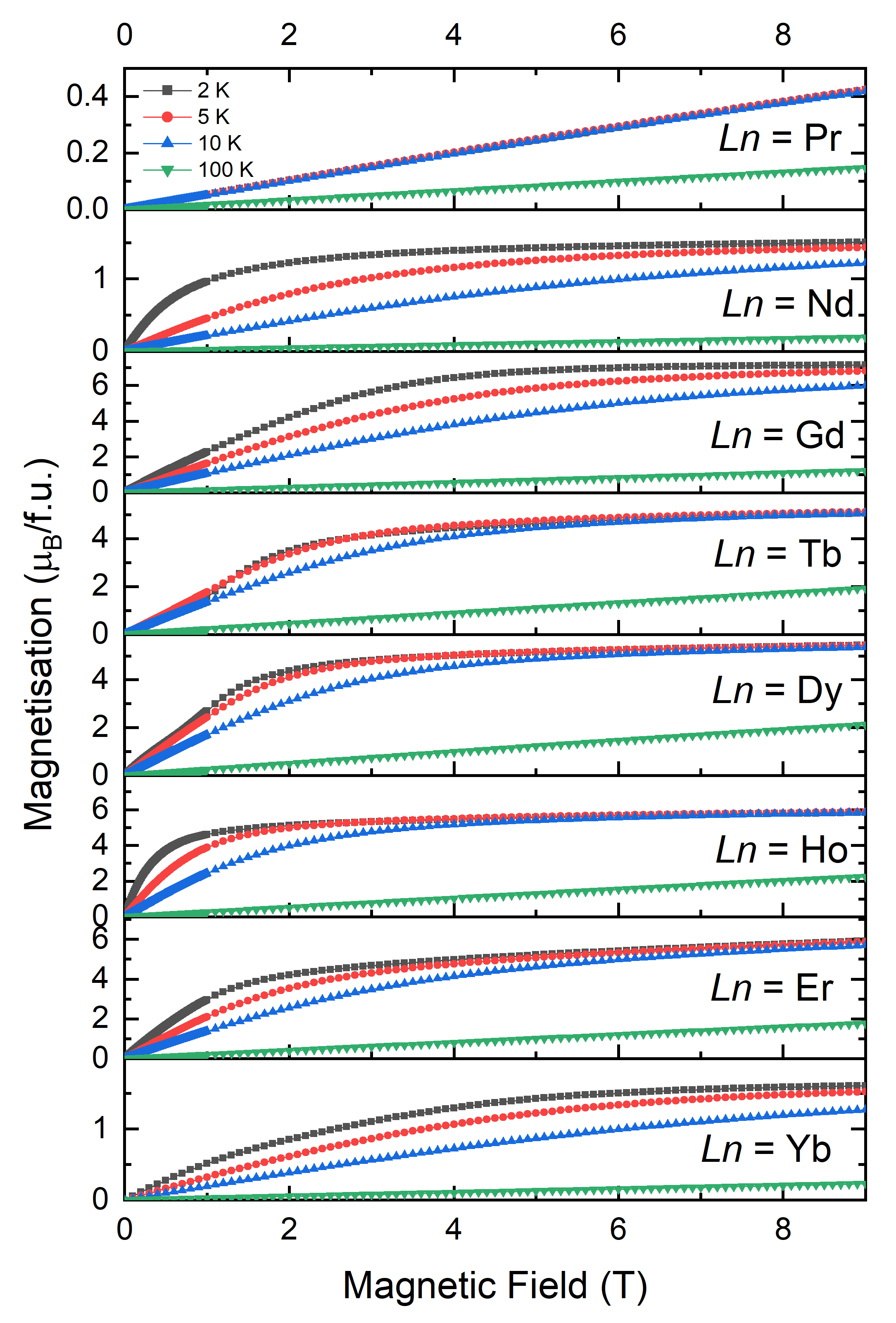}
\caption{Magnetic susceptibility as a function of applied field for the \LnCOB\ samples with \emph{Ln}~= Pr, Nd, Gd, Tb, Dy, Ho, Er and Yb.}
\label{fig:lncobmvh}
\end{figure}

Isothermal magnetisation data for the \LnCOB\ compounds are shown in Fig.~\ref{fig:lncobmvh}. We conclude that \PrCOB\ has a singlet ground state, as observed in other Pr compounds \cite{Zorko2010,Sanders2016,Dun2017a,Mukherjee2017}. The magnetisation saturates at 2 K and 9 T in all other compounds. For an isotropic (Heisenberg) spin system the magnetisation is expected to saturate at a value of $g_JJ$, whereas easy-axis (Ising) systems tend to saturate at half this value, although other contributions may increase $M_{sat}$ above $g_JJ/2$ \cite{Bramwell2000,Dixey2018}. In particular, the exact saturation value for a particular lanthanide ion varies depending on the local point group symmetry (i.e.~CEF) of the atomic site; further experiments such as inelastic neutron scattering are required in order to confirm the anisotropy. The data for \LnCOB\ indicate that \GdCOB\ is likely a Heisenberg spin system, with the remaining compounds showing substantial local single-ion anisotropy. These results are consistent with other lanthanide compounds such as the \emph{Ln}$_3$Sb$_3$Zn$_2$O$_{14}$ kagome lattices \cite{Sanders2016}, titanate pyrochlores \cite{Bramwell2000,Raju1999}, gallium and aluminium garnets \cite{SackvilleHamilton2014,Mukherjee2017a}, hydroxycarbonates \cite{Dixey2018} and metaborates \cite{Mukherjee2017}.

\subsection{Discussion}
The trivalent \emph{Ln}$^{3+}$ ions have highly localised 4$f$ orbitals, meaning that the superexchange between ions -- which depends on the orbital overlap -- is smaller than for the first-row transition metals. We therefore expect superexchange ($J_{nn}$, along the 1D spin chains) to be of a similar magnitude to, or much larger than the dipolar interactions ($D$) depending on the electronic configuration of the lanthanide ion in question, and hence its magnetic moment. The superexchange can be estimated in the mean-field, isotropic approximation using

\begin{equation} 
J_{nn} = \frac{3k_B\theta_{CW}}{2nS(S+1)}
\label{equation:jnn}
\end{equation}

where $S$ in the denominator is the total spin quantum number and $n$ is the number of nearest-neighbour spins (here $n$ = 2) \cite{Ramirez1994}. For systems containing lanthanide ions the spin-orbit coupling cannot be neglected and the quantum number $J=|L \pm S|$ is usually substituted for $S$ in equation~\ref{equation:jnn}, depending on whether the shell is more or less than half-filled. The mean-field approximation is in general a poor model for lanthanide systems due to the strong single-ion anisotropy which is often observed. However, in the absence of inelastic neutron spectroscopy (INS) data the mean-field approximation does allow us to infer some information regarding the magnetic interactions.

In the case of a two-level spin system, it is more appropriate to take $S_{eff} = \frac{1}{2}$. This is true for Nd$^{3+}$, Dy$^{3+}$, Er$^{3+}$ and Yb$^{3+}$, which are Kramers ions with an odd number of $f$-electrons and therefore symmetry-constrained to have an $S_{eff} = \frac{1}{2}$ doublet ground state at low temperatures. Examples may be found in references~\citenum{Ashtar2019,Gao2018,Xing2019,Li2015}. For non-Kramers ions (Tb$^{3+}$, Ho$^{3+}$) this constraint does not apply if the local point group symmetry of the ion is lower than cubic \cite{Walter1984}, as is the case here (\emph{Ln}$^{3+}$ ions in orthorhombic sites). Tb$^{3+}$ and Ho$^{3+}$ have been reported to have $S_{eff} = \frac{1}{2}$ ground states due to mixing of two low-lying singlet states in some compounds\cite{Felsteiner1981,Sibille2017} but this is not the case for \ch{Ho3Mg2Sb3O14} and \ch{Ho3Ga5O12} \cite{Dun2017a,Paddison2019}. In the absence of INS data we cannot confirm the spin anisotropy for the \LnCOB\ compounds with \emph{Ln}~= Nd, Tb, Dy, Ho, Er and Yb, and will therefore calculate the superexchange using both $J=L \pm S$ and $S_{eff} = \frac{1}{2}$.

The dipolar interaction $D$ may be estimated using

\begin{equation} 
D = \frac{-\mu_0\mu^2_{eff}}{4\pi r^3}
\label{equation:dipolar}
\end{equation}

where $r$ is the distance between adjacent \emph{Ln}$^{3+}$ ions in the same or neighbouring chains \cite{Gingras2011}. This is a general expression for $D$ which does not take into account any single-ion anisotropy. Furthermore, a true calculation for $D$ would be long-range and cover many spins. As with equation~\ref{equation:jnn}, however, we here use the general expression to provide a ballpark estimate for $D$ under the stated approximations, since INS data are not yet available.

The sizes of the dipolar and nearest-neighbour exchange interactions were estimated using the magnetic susceptibility data (low-temperature fitting) and the resulting parameters are given in Table~\ref{table:lncobdj}. It has been proposed that the quasi-one-dimensional nature of the magnetism in these compounds is related to the relative sizes of the exchange and dipolar interactions \cite{Mukherjee2017}. Regardless of the size of $J$, we expect to see quasi-1D magnetic behaviour with signatures of low-dimensional ordering at low temperatures (as may be the case for \TbCOB) because $D_{intra-chain}$ is an order of magnitude larger than $D_{inter-chain}$. At this stage we refrain from drawing solid conclusions about the true one-dimensional nature of these compounds, due to the absence of measurements at $T < 2$~K and the potential inadequacy of the mean-field approximations used to derive the interaction energies.

\begin{table*} 
\caption{Dipolar ($D$) and nearest-neighbour exchange ($J_{nn}$) interactions for \LnCOB, \emph{Ln}~= Pr, Nd, Gd, Tb, Dy, Ho, Er and Yb. The negative signs indicate antiferromagnetic interactions.}
\label{table:lncobdj}
\begin{tabular}{c c c c c}
\toprule
\emph{Ln} & $D_{intra-chain}$ (K) & $D_{inter-chain}$ (K) & $J_{nn}$ (K) using $J$ & $J_{nn}$ (K) using $S_{eff} = \frac{1}{2}$\textsuperscript{\emph{a}} \\
\midrule
Pr & --0.36 & --0.03 & --1.38 & N/A \\
Nd & --0.32 & --0.03 & --0.09 & --3.07 \\
Gd & --2.21 & --0.19 & --0.12 & N/A \\
Tb & --2.82 & --0.24 & --0.15 & --8.25 \\
Dy & --3.29 & --0.28 & --0.06 & --4.78 \\
Ho & --3.29 & --0.28 & --0.01 & --0.66 \\
Er & --2.92 & --0.24 & --0.05 & --3.99 \\
Yb & --0.39 & --0.03 & --0.12 & --2.52 \\
\bottomrule
\end{tabular}
\newline
\textsuperscript{\emph{a}}For two-level systems (see text).
\end{table*}

The potential for these compounds to act as magnetocaloric materials was quantified by calculating the change in magnetic entropy per mole, $\Delta S_m$, according to the Maxwell thermodynamic relation \cite{Pecharsky1999}:

\begin{equation}
\Delta S_m = \int_{H_0}^{H_1} \left( \frac{\partial M}{\partial T}\right)_H dH
\label{equation:mce}
\end{equation}

\begin{figure} 
\centering
\includegraphics[width=8.5cm]{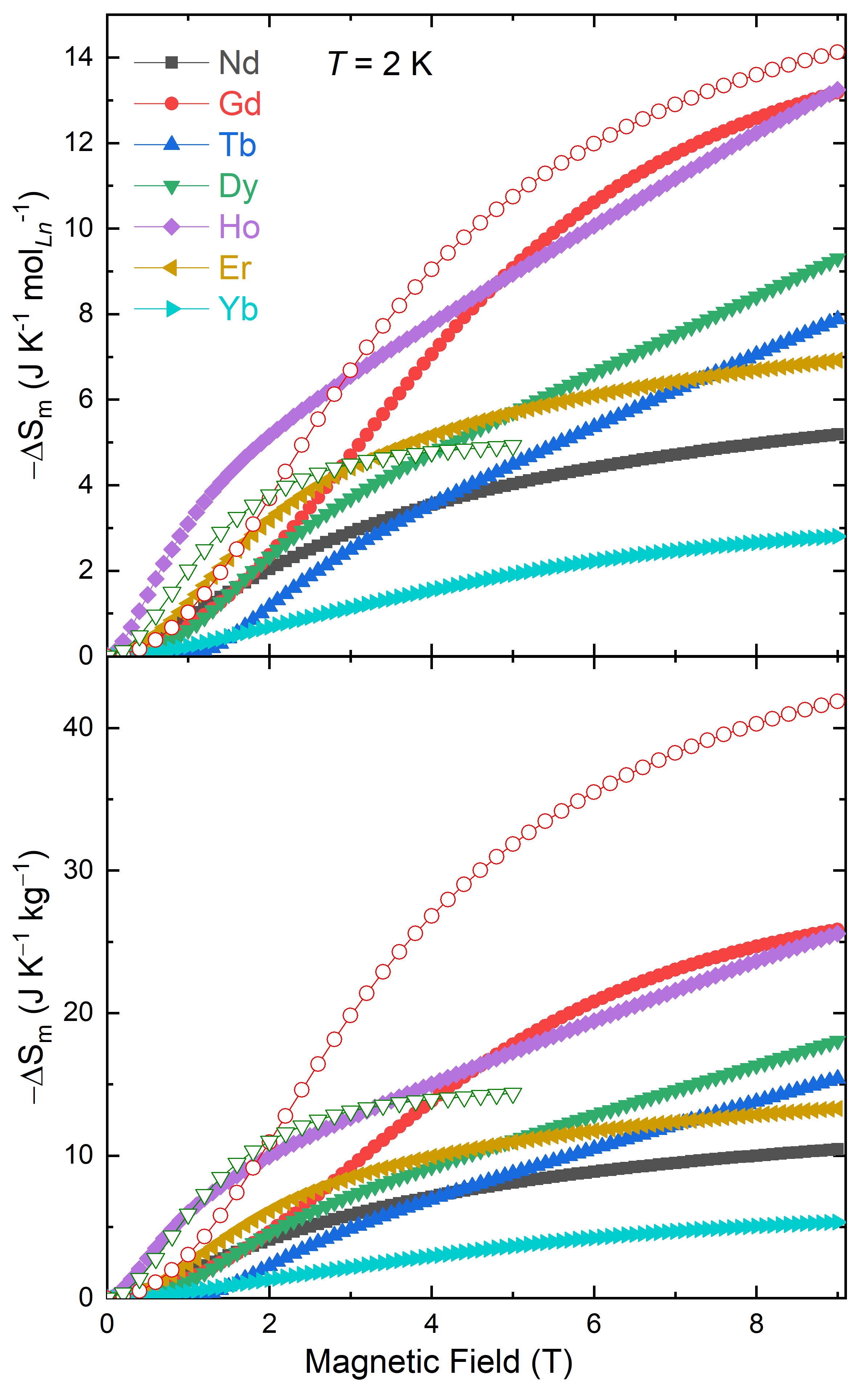} 
\caption{Molar (top) and gravimetric (bottom) magnetocaloric data for \LnCOB\ (\emph{Ln}~= Nd, Gd, Tb, Dy, Ho, Er and Yb; filled symbols) at $T$ = 2 K as a function of applied field, compared with \ch{Gd3Ga5O12} (open circles) and \ch{Dy3Ga5O12} (open triangles) \cite{Mukherjee2018}.}
\label{fig:lncobmce}
\end{figure}

Magnetocaloric data for selected \LnCOB\ samples as a function of applied field are shown in Fig.~\ref{fig:lncobmce}. \GdCOB\ provides the optimal MCE in fields $5 < \mu_0H$(T) $ < 9$, while \HoCOB\ is the best magnetocaloric material in this family at fields below 5~T. Selected data for the \LnCOB\ compounds are compared with the standard magnetocaloric materials \ch{Gd3Ga5O12} (GGG) and \ch{Dy3Ga5O12} (DGG) at low and high fields in Table~\ref{table:lncobmce} \cite{Numazawa2003,Mukherjee2017a}. Here we define a low field as $\mu_0H \leq$~2~T, which is the largest field attainable with a permanent magnet as opposed to a superconducting one. We find that at $\mu_0H = 9$~T, \GdCOB\ is competitive with GGG in terms of MCE per Gd$^{3+}$ ion, but in terms of MCE per kilogram it performs poorly due to the four heavy Ca$^{2+}$ ions per formula unit. At a low field of 2~T, \HoCOB\ has a significantly higher MCE per mole of lanthanide ion than DGG, and a comparable gravimetric MCE.

The straightforward, scaleable solid-state synthesis of \LnCOB, combined with very low ($T < 4$~K) magnetic ordering temperatures, makes these materials attractive candidates for magnetic refrigeration applications. Tuning of the MCE at different temperatures or fields may be possible through partial chemical substitution of the lanthanide ions \cite{Saines2015}. We note particularly that this structure type is highly flexible in allowing substitution of a range of lanthanide ions of different sizes, without any cation mixing that would destroy the quasi-1D magnetic structure. Partial substitution of Sr$^{2+}$ for Ca$^{2+}$ has also been reported, further increasing the family of related compounds \cite{Crossno1997}.

\begin{table*} 
\caption{Comparison of $\Delta S_m$ ($T$ = 2 K) in molar, gravimetric and volumetric units for selected \LnCOB\ compounds and the standard magnetocaloric materials \cite{Mukherjee2017a,Numazawa2003}.}
\label{table:lncobmce}
\resizebox{\textwidth}{!}{  
\begin{tabular}{c c c c c}
\toprule
Compound & Field (T) & $-\Delta S_m$ (J K$^{-1}$ mol$_{Ln}^{-1}$) & $-\Delta S_m$ (J K$^{-1}$ kg$^{-1}$) & $-\Delta S_m$ (mJ K$^{-1}$ cm$^{-3}$)\\
\midrule
\GdCOB\ & 9 & 13.2 & 25.9 & 96.5 \\
\ch{Gd3Ga5O12} & 9 & 14.1 & 41.9 & 296.4 \\
\midrule
\HoCOB\ & 2 & 5.1 & 9.9 & 38.1\\
\ch{Dy3Ga5O12} & 2 & 3.8 & 11.0 & 80.6\\
\bottomrule
\end{tabular}}
\end{table*}

\section{Conclusions}
Twelve compounds in the series \LnCOB\ (\emph{Ln}~= Y, La, Pr, Nd, Sm, Eu, Gd, Tb, Dy, Ho, Er, Yb) have been synthesised using a straightforward and repeatable solid-state procedure. The reported structure has been confirmed using X-ray diffraction. Bulk magnetic characterisation indicates that the quasi-one-dimensional nature of these materials leads to suppression of the magnetic ordering temperatures, which additionally makes these materials good candidates for magnetic refrigeration applications at liquid helium temperatures. These results contribute to the rapidly growing set of lanthanide--alkaline earth borates with low-dimensional structures and novel magnetic properties, such as the Sr$_6$\emph{Ln}Fe(BO$_3$)$_6$, Ba$_3$\emph{Ln}(BO$_3$)$_3$ and \emph{A}Ba\emph{Ln}(BO$_3$)$_2$ (\emph{A}~= Na$^+$, K$^+$, Rb$^+$) structural families \cite{Inoue2016,Gao2018,Sanders2017,Guo2019,Guo2019a,Guo2019b}. Furthermore, the \LnCOB\ structure type is compositionally flexible, which should allow for the realisation of novel low-dimensional magnetic lattices through chemical substitution.

\begin{acknowledgement}
We acknowledge funding from the EPSRC for a PhD studentship and the use of the Advanced Materials Characterisation Suite (EPSRC Strategic Equipment Grant EP/M000524/1). NDK thanks J.~Paddison, P.~Mukherjee and J.~Tuffnell for useful discussions and C.~Liu for assistance in collecting the magnetic data.
\end{acknowledgement}
Data files relating to this article are available at \url{https://doi.org/10.17863/CAM.52921}.
\bibliography{library}

\end{document}